\documentclass[10pt,twocolumn]{article} 
\usepackage{ieee}
\usepackage{times}
\usepackage{graphics}
\usepackage{setspace}
\usepackage{natbib}
\usepackage{fancyheadings}

\newcommand{\postsecskip}{0mm}
\newcommand{\presubsubsecskip}{0mm}
\newcommand{\postsubsubsecskip}{0mm}

\title{Predictable Software --- A Shortcut to Dependable Computing ?}
\author{George Candea \\ \small 
        Computer Systems Lab, Stanford University \\ {\tt \small 
        candea@cs.stanford.edu}}

\begin{document}

\maketitle

\begin{abstract}

Many dependability techniques expect certain behaviors from the
underlying subsystems and fail in chaotic ways if these expectations
are not met.  Under expected circumstances, however, software tends to
work quite well.  This paper suggests that, instead of fixing elusive
bugs or rewriting software, we improve the predictability of
conditions faced by our programs.  This approach might be a cheaper
and faster way to improve dependability of software.  After
identifying some of the common triggers of unpredictability, the paper
describes three engineering principles that hold promise in combating
unpredictability, suggests a way to benchmark predictability, and
outlines a brief research agenda.


\end{abstract}


\section{Introduction}

\vspace{\postsecskip}

Dependability has traditionally been defined as the confluence of
reliability, availability, security, and
safety~\cite{laprie:terminology}; a system is not dependable when any
of these properties is lacking.  Most unplanned violations of
dependability properties trace their roots to some subsystem's
unpredictable behavior in the face of unpredicted stressors.  If the
designer or administrator of a system could accurately reason about
the system's reaction to both known and unknown stimuli, then she
could assess a priori whether that system will satisfy its users'
dependability requirements.  Predictable behavior facilitates accurate
provisioning and enables failure management infrastructures that avert
system-wide failure.

Much deployed software, however, is not predictable.  When given their
expected ranges of input and resource availability, programs generally
output results that conform to their users' expectations.  Unexpected
conditions, however, can lead to cascading failures: an initial fault
perturbs one module's input, which make its output and resource
consumption erratic; this deprives nearby modules of resources and
provides wrong inputs to other modules, which fail and increase the
amplitude of the propagating failure wave.  The ensuing chaos blurs
the cause of failure, making diagnosis progressively more difficult.
A dependable system should never behave unpredictably in the face of
failure; it is OK to fail, but should do so in well-understood,
controlled ways.

Based on the assumption that software works well when not pushed to
unexplored boundaries, this paper posits that system predictability
can be improved by providing programs with the environment they
expect, along with dampening circuits to control propagation of
failure.  Compared to developing new technology for tolerating
failures, improving predictability may provide a cheaper, faster way
to achieve the level of dependability we seek, because it allows
existing techniques to be leveraged effectively.


\section{Unpredictable Behavior}
\label{sec:unpredictable}

\vspace{\postsecskip}

To control unpredictability, we need to first understand the sources
of such unpredictability---they can be internal (e.g., deterministic
bugs), or external.  The proposed approach focuses on external
triggers of unpredictability, so here I survey three categories of
such triggers: unexpected inputs, undue resource utilization, and
unusual failure modes.


\vspace{\presubsubsecskip}
\subsubsection*{Unexpected Inputs}
\vspace{\postsubsubsecskip}

Inputs to a system can be unexpected in at least three ways: in terms
of size, content, and the rate at which input arrives.

{\em Unexpected size}: The vast majority of CERT's 142 alerts
regarding security and availability compromises since year 2000 are
due to buffer overflows, in which input longer than expected
overwrites the program's stack with attacker-provided code.  One of
the most visible cases was SQL~Slammer~\cite{cert}, a
stack-overwriting worm that attacked Microsoft SQL Server's resolution
service, gaining control of the machine and then self-replicating to
other hosts.  Set against SQL~Slammer's estimated impact of more than
\$1 billion, specifying and enforcing the maximum size of each type of
network packet SQL Server accepts seems trivial.

{\em Unexpected content}: A recent exploit in Mailman, the GNU mailing
list manager, allowed attackers to crash it by using a specially
crafted e-mail message that exploited a bug in the message parsing
component~\cite{cve}.  Similar denial-of-service vulnerabilities, this
time related to HTML parsing, have plagued the Apache web server and
the Squid proxy cache~\cite{cert}.  Recently, the most widely used DNS
server (BIND) had a bug that allowed a remote attacker to poison its
name resolution cache~\cite{cve}.  In all of these cases, separately
verifying the validity of the inputs before delivering them to the
vulnerable programs was straightforward, but required thought and,
perhaps, a slight loss in performance.

{\em Unexpected rate of arrival}: Internet services often fail due to
overload conditions.  A notorious example is CNN.com, the online news
provider: On the morning of September 11, 2001, accesses to its front
page increased from 1,400 requests/sec to 3,800 requests/sec within 15
minutes~\cite{lefebvre:cnn_short}.  In spite of CNN.com's load
balancers, dynamic server provisioning systems, and procedures for
reducing HTML complexity for front-page news in response to high
demand, the site collapsed.  Administrators were unable to access the
machines remotely because of thrashing and the number of network
sessions having reached the maximum supported number.  After a few
hours, the site came back up; on that day, CNN.com's estimated peak
load exceeded 30,000 requests/sec.  The system would have been able to
handle this peak load, but was not able to adapt fast enough when
workload was doubling every 5 minutes; suitable admission control
would have allowed CNN.com to stay within operating range.  Although
large Internet services have several mechanisms in place to control
the magnitude of request load, few (if any) check for how the rate of
arrival evolves.

The inputs described in this section originated at sources that were
outside the receiver's administrative realm (e.g., end users, remote
systems).  The next section addresses another category of input: the
output of software modules that the receiver directly interacts with,
within the same realm of control.


\vspace{\presubsubsecskip}
\subsubsection*{Unaudited Output}
\vspace{\postsubsubsecskip}

In order to manage behavior of large programs, software engineers
developed the concept of modules with well-defined interfaces, which
isolate code within a known set of behavior boundaries.  Over time,
however, module behavior changes in subtle ways as internal code gets
modified and optimized.  Even if behavior did not change, programmers
use modules in ways that the modules' original designers had not
envisioned, exercising poorly tested paths that lead to unexpected
behavior.  Programmer churn compounds these dynamics, because new
programmers modify and/or use code they do not understand.  In a study
of a high-end IBM operating system~\cite{sullivan:defects}, code reuse
has been given as one explanation for the high incidence (56\%) of
boundary condition bugs that had a high impact on the system's user
population.

Misbehaving modules' outputs become other modules' inputs and can lead
to unexpected effects outside the system.  A recent survey of email
administrators~\cite{kakes:email_short} found that botched upgrades
and misconfiguration caused 42\% of all lost email incidents.
Similarly, downtime in the US public telephone network increased from
15 million customer-minutes per month in
1992-1994~\cite{kuhn:humanerr} to 155 million per month in
2000~\cite{enriquez:pstn_short} --- one explanation is the widespread
Y2K patches.  If outputs could be kept conformant independently of
patches, misconfiguration, and misuse, then software evolution would
pose less of a risk.


\vspace{\presubsubsecskip}
\subsubsection*{Reckless Resource Usage}
\vspace{\postsubsubsecskip}

Software aging is the process by which the state of long-running
programs degrades over time through exhaustion of system resources,
data corruption, or numerical error accumulation.  Aging may lead to a
potpourri of unexpected behavior: performance degradation, crashes,
hangs, etc.  Exhausted resources can range from memory and disk space
to CPU time and operating system structures like file descriptors.
Software aging is particularly problematic for popular Internet
services, because their resource consumption is workload-driven and
they face large workloads on a regular basis; scientific applications
suffer as well, because they run for a long time.  A recent version of
Apache leaks 80 bytes of memory every time an HTTP request has a
linefeed on a line by itself (which is legal input); bombarded with
large chunks of linefeeds, Apache runs out of memory and crashes
within seconds~\cite{cve}.  Software tends to become increasingly less
predictable when faced with a shortage of resources, both because it
may not check for such conditions and because code paths that are
normally not exercised now get executed.  In the Apache example, on
Linux kernel 2.4.20, once the leaking process' footprint exceeds the
amount of swap space, a kernel bug in the swap code freezes the entire
machine in about 50\% of the cases.

Many server applications preallocate memory pools and do their own
memory management.  This suggests that, if there is a chance memory
may be insufficient, they insist on knowing ahead of
time---predictable resource availability is sometimes more important
than resource availability itself.


\bigskip

Historically, simplicity has been the key to dependability, because
simplicity begets predictability; simple systems tend to work, while
complex ones don't.  The challenge is to build complex systems that
are dependable; one way to do so is by preserving the simplicity of
the interactions between their components.  Having motivated the need
for predictability, I will now argue for putting order in the
seemingly anarchic universe of software.


\section{The Physics of Software}
\label{sec:laws}

\vspace{\postsecskip}

Mature engineering disciplines are constrained and, at the same time,
blessed with the immutable laws of physics.  They take the form of
macroscopic, descriptive laws that capture physical invariants (e.g.,
Joule's heat dissipation law for electric circuits $P=RI^2$).  These
are safe-to-make assumptions that engineers use to reason about the
future behavior of an electric circuit, a steel structure, or a
chemical process; no effort is required to preserve such invariants.


This paper argues that we restrict the inputs, outputs, and internal
states of software, the same way laws of physics constrain physical
systems.  Unlike the physical world, however, in software we need to
formulate and enforce such invariants ourselves.  The rest of this
section describes three ways to define such invariants; although
described with respect to individual modules, the three methods apply
hierarchically at all levels of granularity within a system, from
small components to large subsystems.


\vspace{\presubsubsecskip}
\subsubsection*{Fail Early}
\label{sec:inputs}
\vspace{\postsubsubsecskip}

It is acceptable for a program to fail when given inputs that the
designer had assumed impossible, but it is not acceptable for such
inputs to be allowed.  Electric circuits, for instance, malfunction if
exposed to current exceeding their design point; their inputs,
however, are controlled by fuses---wires that physically disintegrate
when carrying current above a certain threshold.  The absence of fuses
and circuit breakers would make household electricity and many other
applications impractical, due to the fire danger and other mayhem
resulting from overloaded circuits.  Laws of physics cannot prevent
all power spikes and miswiring, but they can guarantee that excessive
current will not flow through a fuse.

A {\em software fuse} is a filter that drops any input that does not
meet the receiving program's expectations.  The input invariants
encoded in fuses need to be explicit about at least the bounds on the
length of acceptable input, content (e.g., only accept ASCII
characters), and nature of workload---in effect stating acceptability
properties~\cite{rinard:acceptability} for inputs.  Stateful
packet-inspecting firewalls are fuses: they enforce fairly
sophisticated invariants on the kind of network traffic that passes
through, such as blocking non-HTTP communication tunneled over the
HTTP permitted by corporate firewalls.  Input invariants do not have
to be static: dynamic feedback loops between the system and its input
fuses can parameterize adaptive invariants.  Analogously to band-pass
filters in analog circuits, input invariants should limit the rate at
which workload varies, to prevent the kind of failure CNN.com
experienced on Sep. 11, 2001.  Input fuses can span inputs of multiple
modules, to enforce higher granularity properties.  Given the
frequency with which systems fail due to bad inputs or overload, one
can only conclude that programmers are pathological optimists.



\vspace{\presubsubsecskip}
\subsubsection*{Kill the Gluttons}
\label{sec:resources}
\vspace{\postsubsubsecskip}

A large fraction of downtime results from undue resource utilization,
which deprives other modules of the resources they need; such
resources consist of bits (memory, disk, network bandwidth, etc.) and
time (CPU time, transaction commit time, etc.).  Running out of
resources is a normal occurrence and dependable applications must
handle it.  The fact that memory leaks are still rampant in
garbage-collected J2EE applications shows that language constructs are
not sufficient for taming resource unpredictability---instead, a
congruency between the resource model and underlying reality is
required.  Exporting accurate resource models to application
programmers without requiring them to perform explicit management
(e.g., via {\tt malloc} and {\tt free}) is challenging, but necessary
for predictability.  Flexible resource models, such as
leases~\cite{gray:leases_short} and market-based node
management~\cite{coleman:oncall_short}, can help, if their paradigm is
assimilated in the programming languages or framework.

Once the constraints of the resource model are properly exposed in the
form of known invariants, any violation can legitimately be considered
evidence of a bug.  {\em Resource cops} in the underlying execution
framework can then promptly suspend or terminate the offender.  The
simplest form of resource control allows a system to execute until it
runs out of some resource, or its utilization exceeds a threshold, and
then reboots it.  In multi-tiered systems, resource utilization can be
tied to requests and each incoming request be given a time-to-live;
should the time-to-live reach zero during processing, the request gets
squashed and all associated resources are freed, thus coercing
resource gluttony into a clean request failure.  A watchdog is a good
example of a resource cop: in PHP, a server-side scripting language,
timers will terminate a script when it exceeds its allotted time, thus
freeing the CPU and preventing further resource hogging.

\vspace{\presubsubsecskip}
\subsubsection*{Fail in Known Ways}
\label{sec:failures}
\vspace{\postsubsubsecskip}

When a system fails, either it provides strange results (Byzantine
behavior), or becomes disabled and provides no answers until it is
recovered (fail-stop behavior).  A dependable system never exhibits
Byzantine behavior and only fails into one of a small set of states,
thus making recovery easier and halting further propagation.


Verifying that the output of a program has a desired property is often
easier than producing output with that property (e.g., prime
factorization of an integer).  Programs can therefore have simple,
orthogonal {\em output guards} that capture the invariants required of
correct output, to as fine a granularity as is practical.  Since
verifying properties of a data set often requires different algorithms
than producing the data set (e.g., sorting a list), output guards can
legitimately increase confidence in the correctness of output, unlike
N-version programming.  For example, the U.S. National Security Agency
locks up uncertified operating systems and applications inside virtual
machines and monitors their network communication to ensure that the
application inside does not violate desired security
invariants~\cite{rosenblum:nsa}.  When an output guard detects a
violation, it can suspend or stop the module, thus coercing Byzantine
into fail-stop behavior.

Output guards can enforce larger granularity constraints by spanning
the output of multiple modules.  Should enforcement of desired
invariants become too expensive, output guards can sample the output,
rather than check every single response.  Sophisticated guards can
enforce output invariants by correlating outputs to the inputs that
generated them, as suggested in~\cite{rinard:acceptability}.

Impending failure can sometimes be inferred by monitoring system
aspects that are not related to output.  For instance, the earlier PHP
watchdog timer example shows how hang failures can be coerced into
crash failures.  Another example maps observed low-level faults onto a
set of faults known to be well tolerated by existing recovery
code~\cite{nagaraja:fme}.  Checking data structures with {\tt
assert()}-like macros can turn runtime violation of data invariants
into a suspend or reboot of the faulty module, thus preempting a wider
disaster.  In most systems, data structure inconsistencies are often
indicators of much deeper problems, that are easier to deal with by
stopping; limping along might further compromise system integrity and
uselessly consume resources.


\section{From Predictability to Dependability}
\label{sec:discussion}

\vspace{\postsecskip}


Immediate detection of invariant violations in inputs, outputs, and
resource usage enables fast fault detection, preventing further system
failure.  Admission control tuned to the CNN.com cluster's ability to
scale under load could have avoided its collapse on Sep. 11.  Instead
of fixing insidious bugs, one might simply prevent the bug-triggering
inputs from entering the system.  Most of the bug fixes addressing the
SQL~Slammer, Mailman, Apache, Squid, and BIND exploits (described in
Section~\ref{sec:unpredictable}) took the form of more thorough input
checking; the same effect could have been obtained out in the field by
operators interposing input fuses.

Output guards and input fuses can dampen the domino effect of a
propagating fault through containment.  This is particularly important
in large scale enterprise systems, where a major source of failures
results from poor integration of new code with legacy software.
Connection frameworks, such as JCA (Java Connector
Architecture~\cite{jca}), are a good vehicle for implementing fuses
and guards, that turn poorly understood behavior of legacy components
into predictable behavior.  Resource cops can reside inside J2EE
application servers, which are frequently used for legacy integration.
Engineering away unpredictability with guards, fuses, and resource
cops is often easier than rewriting old software (which may sometimes
not even be an option), and hence cheaper.  Enforced predictability
could also encourage more frequent reuse of debugged software modules.

Prompt fault detection plus fast recovery is the recipe for high
availability.  Well-understood failure states enable the development
of fast and effective recovery procedures; the success of the
transaction model~\cite{gray:transproc} is testimony to how a simple,
well-defined fault model can improve the dependability of
applications.

The threat of program termination when outputs or resource consumption
do not meet expectations should motivate developers to write correct
and complete input fuses for their modules.  The mere formulation of
input invariants improves a developers' understanding of the system
and helps find bugs in the module's logic.  Guards and fuses that
catch wrong assumptions also provide immediate feedback and facilitate
debugging during development.


Rigorously developing fuses and guards can improve the effectiveness
of testing, thus increasing software quality.  By excising the task of
input and output validation from the main code into separate fuses and
guards, we essentially develop live testing modules.  Since they are
considerably simpler, their correctness could be verified using formal
methods (in fact, their simplicity automatically makes them less prone
to bugs).  Then, the system itself need only be tested on filtered
inputs, which can be considerably fewer than unfiltered ones,
depending on the application.


Predictably faulty behavior can be efficiently compensated for, thus
turning expensive unplanned downtime of buggy software into planned
downtime.  For instance, a web server that leaks memory at a constant
rate can be rebooted shortly before its predicted failure point.
Unpredictability calls for overprovisioning to accommodate variations,
which means more resources to be purchased and managed; predictable
systems may allow for lower error margins and require less human
oversight.  For example, in the US public telephone system, 30\% of
failures are due to hardware, yet hardware causes only 19\% of total
downtime~\cite{enriquez:pstn_short}.  Switching hardware has simple
failure semantics and replacement procedures, enabling technician
crews to recover it quickly.


Predictable behavior can reduce the chances for operator errors and
improve a system's perceived availability.  Various psychological
studies have shown that, when acting under pressure, humans are poor
at analyzing the situation and making intelligent
decisions~\cite{reason}; most human operator errors result from such
hasty actions.  Predictable failure behavior offers opportunities to
accumulate experience with similar failures through repetition, as
well as to develop automated procedures for failure management.
Predictability may also reduce the number of system ``knobs,'' which
means less risk for error.  Additionally, end user perception of the
system's dependability could increase with predictability; a study of
computer users has found that increased predictability of terminal
response time gave rise to more productive and satisfied users than
unpredictable access, even when response time was as high as 5
seconds~\cite{miller:variability}.



\section{Variance as a Predictability Benchmark}
\label{sec:variance}

\vspace{\postsecskip}

Improving predictability may take a toll on some system properties
(e.g., performance), and reward the designer with improvements in
other areas (e.g., recovery time).  Choosing the right combination of
predictability, performance, cost, etc. requires a way to measure
changes in predictability.  Sometimes predictability is not important,
while in other cases it may yield high value.  For example, in a
triple-modular redundant design, the system mean-time-to-failure MTTF
is reduced, because the system stops when any one of its 3 modules
fails, but mean-time-to-recover MTTR from a single faulty value is
predictably zero.

The definition of ``right tradeoff'' generally relates to user
expectations or cost.  It has been argued that reducing MTTR improves
availability if MTTF stays the same, because availability is
MTTF/(MTTF+MTTR)~\cite{patterson:roc_tr_short}.  However, a small MTTR
does not guarantee fast recovery, if the TTR distribution is
heavy-tailed.  Failures under high load usually lead to the longest
recovery times, thus affecting even more users, so a lower-variance
TTR distribution might actually be more useful, even if it entails a
higher mean.  A web-based service with an end-to-end \{MTTR, standard
deviation\} of \{$\mu$=6~sec,~$\sigma$=2.5\} will have better end-user
perceived availability if it reduces variance to zero, even if that
requires raising MTTR: \{$\mu$=8~sec,~$\sigma$=0\}.  In the latter
case, an observed outage will never last more than 8 sec (which is
known to be the patience threshold for humans using web-based
services~\cite{bhatti:response}).  The former system ($\mu$=6~sec)
allows a significant fraction of outages to last longer than 8
seconds, or allows a few outages to take excessive amounts of time.

To benchmark a system's predictability, we would first need to
establish an invariant and measure how close (heuristically) to that
invariant the system's behavior is.  Consider measuring the
predictability of recovery in an Oracle 9i database system: set the
{\tt FAST\_START\_MTTR\_TARGET}
parameter~\cite{lahiri:fast_restart_short} to $t=1~\mathrm{minute}$
(representing a recovery time invariant).  Then place the standard
TPC-C load on the system, crash it, and measure recovery time.  Repeat
several times and compute the variance.  Repeat the same experiment
for other values of $t$.  The predictability metric tells us how
tightly clustered the recovery times are around the chosen {\tt
FAST\_START\_MTTR\_TARGET} value.  A low variance can offer confidence
that, in production, the DBMS will provide recovery times within the
predicted bounds.


\section{Research Agenda}
\label{sec:future}

\vspace{\postsecskip}


The most common approach to making programs work as expected consists
of finding and fixing bugs in the source code.  External invariant
enforcers can treat programs as black boxes, but they are likely to be
less effective against internal triggers of unpredictability than
against external ones.  This research agenda will focus on bringing
predictability to {\em existing} software modules that {\em must} be
treated as black boxes.  The overhead of creating fuses, guards, or
cops is best justified when modification of the software itself is
difficult or impractical, thus making legacy software a prime
candidate.  Systems connected to the open Internet are particularly
interesting due to the unpredictability of their workload.

{\em Smart fuses in C++ and Java:} The easiest software fuse is an
{\tt if} statement or a simple admission control rule; the more
sophisticated invariants, however, require expressive declarative
languages that can capture temporal properties.  Unfortunately,
programmers have enough trouble mastering their main programming
language, so mandating additional languages with substantially
different paradigms is impractical.  Two artifacts that can be of real
help to practitioners would be a set of parameterized fuse prototypes
(e.g., that enforce tunable temporal properties of workload evolution)
and libraries that allow programmers to write their own fuses,
directly encoding input invariants.  Model-checking and other formal
methods can be used to increase confidence in the correctness of the
fuse libraries.

{\em Aids for invariants:} Success of fuses, guards, and cops depends
on the correctness and completeness of the invariants they enforce.
Not only do common programmers not have good tools to formulate such
invariants, but they don't always realize all the assumptions they're
making.  Recent work~\cite{ernst:daikon,demsky:roles} has made
important progress in learning invariants by observing program
behavior, but more remains to be done for these techniques to work
with legacy black boxes.  Deduced invariants could be presented to
programmers, or enforced and verified on subsequent executions of the
programs.  Such tools could also be used standalone by developers to
better understand the system they are assembling.

{\em Nontrivial output guards:} Substantially meaningful output
invariants can most likely not be stated independently of the inputs
generating that output.  As such, the more sophisticated the output
invariant, the more likely it becomes that the guard duplicates parts
of the program itself.  The solution lies in identifying abstract
classes of modules that can share guard constructs (e.g., all Java
EJBs support standard sets of methods that can be guarded by common
invariants), and then using output sampling to statistically enforce
the output invariants.  Identifying optimality of when and how to
sample output is another interesting challenge.

{\em Reasonable resource models:} Enforcing invariants on resource
usage without making things worse could prove a challenge: resource
limits tight enough to be useful might be too tight, and limits that
avoid problems might be too loose.  Global resource optimization
problems have been studied at length in operating systems, but under a
fairly restricted resource model.  Exposing more flexible models could
yield more efficient resource control.  For example, exclusive use of
leased resources~\cite{gray:leases_short} in a prototype J2EE
application server could enable zero-downtime prophylactic rebooting
of components to combat software aging.

{\em System macrocompiler:} Type safety has gone a long way in
validating inputs and outputs of programs.  Compilers have improved
low-level predictability and correctness, as well as given us a way to
reason about programs.  Unfortunately, high level properties of
heterogeneous systems cannot be verified by compilers.  A macrocompiler
would validate compatibility between assemblies of black boxes and the
corresponding invariants captured by their fuses, guards, and resource
cops.  A macrocompiler might also enable safe composition of higher
level invariants out of properties enforced at lower levels, such as
type safety. Component properties, enforced by guards, could then be
used to reason about emergent properties.  A macrocompiler using
aspect-oriented approaches~\cite{aop} can automatically interpose
fuses and guards on arbitrary classes of objects in heterogeneous
environments ranging from C/C++ to Java and J2EE.

{\em Metrics and deployment:} Dependability benchmarking is still an
elusive goal, but predictability metrics can bring us closer to a
solution.  I've suggested experimental variance as a measure of
predictability, but better, more analytical techniques are needed.
Quantifying the tradeoffs involved in engineering predictability could
also help with identifying the parts of a system that would most
benefit from improved predictability; this would enable the
incremental retrofitting of predictability into existing systems in a
cost-effective way.


\section{Summary}
\vspace{\postsecskip}

Software fuses and resource cops keep software within its comfort
zone; if bugs continue to manifest, then output guards and other
resource cops bring erratic behavior into compliance, transforming a
black-box program into a fail-stop module.  If programs behave better
when external conditions match those envisioned by its designer, then
such predictability begets dependability.  Some types of fuses,
guards, and cops are already in ad hoc use today; building smarter
ones and systematically connecting them to programs can improve
predictability of software.  The proposed approach may be a productive
alternative to fixing pernicious bugs or rewriting flakey software,
and may ease the integration of imperfect software into complex
systems.


\section{Acknowledgements}

I am grateful to Timothy Chou for planting the seed that led to the
ideas described here.  Many of the practical aspects of the proposal
trace their roots to discussions with my doctoral advisor, Armando
Fox, and the ROC group.  I am indebted to Katerina Argyraki, Aaron
Brown, David Cheriton, Steve Gribble, Kim Keeton, Bill Joy, Martin
Rinard, Bjarne Stroustrup, and my colleagues in SWIG for their
invaluable help with finetuning the ideas and their presentation.


\setlength{\bibsep}{0mm}
\setlength{\bibhang}{5mm}

\newpage
\small
\bibliographystyle{abbrv}
\vspace{-1mm}
\bibliography{../rr.bib}

\end{document}